\documentclass[aps,prl,epsfig,twocolumn,showpacs]{revtex4}
\usepackage{graphicx}
\usepackage{dcolumn}
\usepackage{bm}
\usepackage{psfig}
\usepackage{amsmath}
\usepackage{amssymb}

\newcommand{\ket}[1]{|#1\rangle}
\newcommand{\bra}[1]{\langle#1|}

\newcommand{\ncd}{\newcommand}
\ncd{\QC}{$\mbox{QC}_{\cal{C}}$}
\ncd{\QCpr}{${\mbox{QC}_{\cal{C}}}^\prime\;$}
\ncd{\QCns}{$\mbox{QC}_{\cal{C}}$}
\ncd{\QCprns}{${\mbox{QC}_{\cal{C}}}^\prime$}
\ncd{\cskN}{{|\phi_{\{\kappa\} } \rangle}_{{\cal{C}}_N}}
\ncd{\cskNpr}{{|\phi_{\{\kappa^\prime\} } \rangle}_{{\cal{C}}_N}}
\ncd{\cskNtil}{{|\phi_{\{\tilde{\kappa} \} }
\rangle}_{{\cal{C}}_N}}
\ncd{\csk}{{|\phi_{\{\kappa\} }
\rangle}_{\cal{C}}}
\ncd{\csktil}{{|\phi_{\{\tilde{\kappa} \} }
\rangle}_{\cal{C}}}
\ncd{\cskf}{|\phi_{\{\kappa\} }
\rangle_{\cal{C}}}
\ncd{\csktilf}{|\phi_{\{\tilde{\kappa} \} }
\rangle_{\cal{C}}}
\ncd{\bracsk}{\mbox{}_{\cal{C}}\langle\phi_{\{\kappa\} }|}
\ncd{\bracsktil}{\mbox{}_{\cal{C}}\langle\phi_{\{\tilde{\kappa} \}
}|} \ncd{\nbracsk}{\mbox{}_{\cal{C}}\langle\phi_{\{\kappa\} }}
\ncd{\nbracsktil}{\mbox{}_{\cal{C}}\langle\phi_{\{\tilde{\kappa}
\} }} \ncd{\cs}{|\phi \rangle_{\cal{C}}\;} \ncd{\csns}{|\phi
\rangle_{\cal{C}}}
\ncd{\nbgh}{\text{nbgh}} \ncd{\Sab}{S^{ab}}
\ncd{\Sba}{S^{ba}}
\ncd{\ds}{\displaystyle} \ncd{\ovl}{\overline}

\begin{document}
\setlength{\hoffset}{-0.2 cm}


\title{Toward a more economical cluster state quantum computation}
\author{M. S. Tame$^1$, M. Paternostro$^1$, M. S. Kim$^1$, V. Vedral$^{2,3,4}$}
\affiliation{$^1$School of Mathematics and Physics, The Queen's University, Belfast, BT7 1NN, UK\\
$^2$Institute of Experimental Physics, University of Vienna, Boltzmanngasse 5, 1090 Vienna, Austria\\
$^3$The Erwin Schr\"{o}dinger Institute for Mathematical Physics,  Boltzmanngasse 9, 1090 Vienna, Austria\\
$^4$The School of Physics and Astronomy, University of Leeds, Leeds, LS2 9JT, UK}

\date{\today}
\begin{abstract}
We assess the effects of an intrinsic model for imperfections in cluster states by introducing {\it noisy cluster states} and characterizing their role in the one-way model for quantum computation. The action of individual dephasing channels on cluster qubits is also studied. We show that the effect of non-idealities is limited by using small clusters, which requires compact schemes for computation. In light of this, we address an experimentally realizable four-qubit linear cluster which simulates a controlled-{\sf NOT} ({\sf CNOT}).
\end{abstract}

\maketitle

 
The one-way quantum computer is a model for quantum computation (QC) exploiting multipartite entangled resources, the {\it cluster states}, and adaptive single-qubit measurements~\cite{rb0}. Quantum information processing (QIP)
is performed in this model using a large enough cluster state and appropriate measurements~\cite{RBH}. The basic features of one-way QC have very recently been experimentally demonstrated in~\cite{vlatko}. Despite the stimulating possibilities offered by this model, cluster state QC is still an open field of research from both a theoretical and practical viewpoint. On one hand, we have witnessed the first attempts of combining cluster state and linear optics in order to increase the efficiency of existing all-optical QC schemes~\cite{nielsenprl}. On the other hand, it is important to develop the model by studying the effects of imperfections on the performances of one-way QC. This is the central aim of this work. So far, it is not clear if the cluster state model, exposed to sources of noise, would perform better than the standard quantum circuit model. Very recently, there have been some intriguing investigations in this sense~\cite{longnielsen}. We introduce a realistic model for imperfect cluster state generation and study how this affects both the intrinsic properties of cluster states and the basic ingredients in computation. We also address the effect of environmental decoherence due to dephasing channels individually affecting the qubits in one- and two-dimensional configurations. In this context, 
an informative initial step
is the study 
of the state fidelity of cluster states~\cite{durbriegel}. We show that in order to limit the spoiling effects of non-idealities, one must deal with small clusters of just a few qubits. This imposes severe contraints on the use of this model and paves the way toward research of more compact protocols for gate simulation and QIP. We find such a possibility, addressing a $\sf CNOT$ simulated through a simple four-qubit linear cluster and outline an all-optical setup where it can be implemented. Our study contributes to the analysis of more realistic one-way QIP and highlights the existence of economical schemes for cluster based QC. Our proposal overcomes the {\it expensive} schemes for gate simulation proposed so far~\cite{RBH} and significantly reduces the requirements for experimental implementations.

{\it The model}- Given a set of qubits occupying the sites of a square lattice ${\cal{C}}$, a cluster state $\ket{\phi_{ \{ \kappa \} }}_{\cal C}$ is a pure entangled state characterized by the eigenvalue equations $K^{(a)}\ket{\phi_{ \{ \kappa \} }}_{{\cal C}}=(-1)^{\kappa_a}\ket{\phi_{ \{ \kappa \} }}_{{\cal C}}$. Here $K^{(a)}=\sigma_x^{(a)}\otimes_{b}\sigma_z^{(b \in \nbgh(a)\cap{\cal{C}})}$ are {\it correlation operators} forming a complete set of $|{\cal{C}}|$ commuting observables for the qubits in ${\cal{C}}$~\cite{RBH}. The set $\{ \kappa \}$, which completely specifies the cluster state $\ket{\phi_{ \{ \kappa \} }}_{\cal C}$, contains the values $\kappa_{a}\in\{0,1\}~\forall{a}\in{\cal C}$, while $\sigma_{x,z}$ are the $x$ and $z$-Pauli matrices respectively. 
Cluster states $|\phi \rangle_{{\cal{C}}}$ with $\kappa_{a}=0~(\forall{a}\in{\cal C})$ are generated by preparing 
a register of qubits within ${\cal C}$ in the state $\otimes_{a\in {\cal{C}}} |+\rangle_{a}$, where $\ket{\pm}_{a}=(1/\sqrt{2})(\ket{0}\pm\ket{1})_{a}$,
and applying the transformation $S^{({\cal{C}})}=\prod_{a,b\in{\cal{C}}|b-a\in
    \gamma_d} \Sab$
with $\gamma_1=\{1\},\,\gamma_2=\{(1,0)^T,(0,1)^T\}$ and
$\gamma_3=\{(1,0,0)^T,(0,1,0)^T, (0,0,1)^T \}$ for the respective dimension of the cluster. 
Each $\Sab$ is a controlled $\pi$-phase gate
$\Sab= |0 \rangle_a \langle 0| \otimes {\openone}^{(b)} + |1 \rangle_a \langle 1| \otimes \sigma_z^{(b)}$.

However the realization of the $S^{ab}$ gates can be inherently imperfect. For example, an optical-lattice loaded by a bunch of neutral atoms, where the two-qubit interactions are via controlled collisions, has been suggested to embody a one-way QC~\cite{bloch}. The efficiency of the gates depends on the control over the strength of the interactions (which can fluctuate over the sites of the lattice, generating inhomogeneities) and the interaction time. These parameters fix the amount of the controlled phase imposed by each interaction. In addition, the initial filling-fraction per site in the lattice influences the performances of the gates~\cite{bloch}. If the degree of control is not optimal (which may be the case when a large number of qubits is considered), imperfect (inhomogeneous) interactions throughout the physical lattice are settled. This can be addressed by considering the imperfect operations 
\begin{equation}
    \label{Sabdefdirty}
    \Sab_D= |0 \rangle_a \langle 0| \otimes {\openone}^{(b)} + |1 \rangle_a \langle 1| \otimes \left(|0 \rangle_b \langle 0|-e^{i \theta_{a}}|1 \rangle_b \langle
    1| \right),
\end{equation}
which add unwanted phases $\theta_{a}$ to the desired value $\pi$. Applied to the initial state of a $N$-qubit register (prepared via the idealized protocol), they generate a {\it noisy cluster state} $\ket{\phi}^{D}_{{\cal C}_{N}}$. The use of this set of operations profoundly modifies the structure and properties of a $N$-qubit cluster with respect to the ideal $\ket{\phi}_{{\cal C}_N}$. Explicitly, a noisy linear cluster state becomes
$|\phi \rangle_{{\cal{C}}_N}^D={2^{-{N}/{2}}}\sum_{{\bf z_i}}\prod^{N-1}_{j=1}(-e^{i\theta_j})^{z_j  z_{j+1}}|{\bf z_i}\rangle$,
where $z_j$ is the value of the $j$-th binary digit of the integer ${\bf z_i}$ ($i\in[0,2^{N}-1]$).

\begin{figure}[b]
{\bf (a)}\hskip4cm{\bf (b)}
\centerline{\psfig{figure=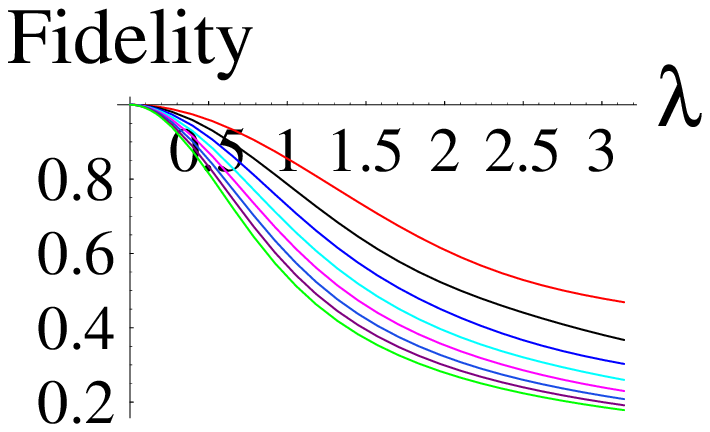,width=3.7cm,height=2.7cm}\hspace*{0.7cm}\psfig{figure=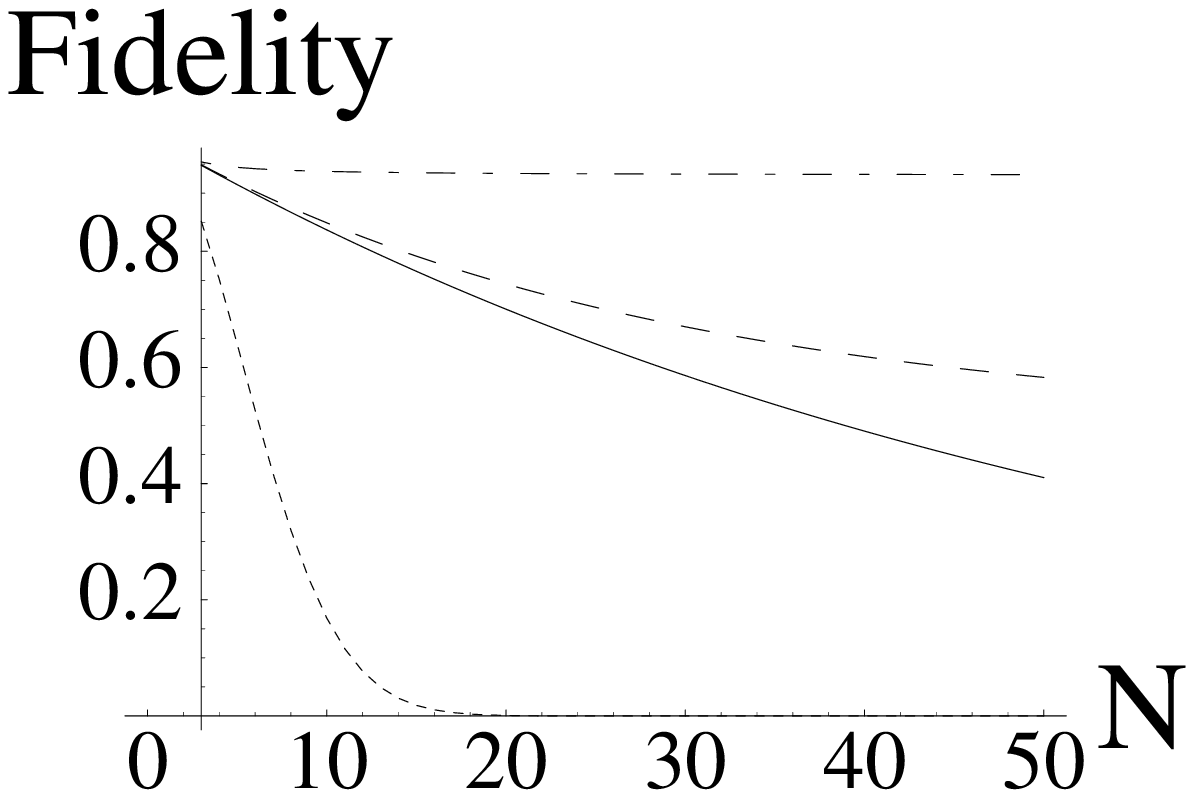,width=4.1cm,height=2.8cm}}
\caption{{\bf(a)}: $\bar{F}_{N}$ against $\vert{\bf r}_{j}\vert = \lambda$ for $p(\theta_j)=\lambda^{-1}$. From top to bottom $N=3\rightarrow{10}$; {\bf (b)}: Dephasing effects on multiqubit states. The curves from top to bottom represent $N$-qubit $W$, $GHZ$, linear cluster and $N \times N$ cluster states. For convenience, the rescaled dephasing time is $\Gamma{}=0.062$ (corresponding to $F^{C_{lin}}_{25}=0.5$). Similar behaviors are found for different values of $\Gamma{}$.}
\label{fig:fig2}
\end{figure}

In order to characterize the effect of this kind of non-ideality, an immediate benchmark is provided by the fidelity between ideal and noisy cluster states. The overlap $f_{N}=\,^{}_{C_N}\langle \phi |\phi \rangle^D_{C_{N}} $ leads to
$f_{N}=2^{-N}\sum_{{\bf z_i}}\prod^{N-1}_{j=1}e^{i \theta_jz_jz_{j+1}}$.
There are $2^N$ terms in this expression, with the fidelity $F_N=|f_N|^2$. 
However, our assumption is that the control on the phases introduced by the qubit-qubit interaction is only limited. Thus,  there is a lack of knowledge about the values of $\theta_{j}$'s in a noisy cluster state, which means each of them must be averaged over an appropriate probability distribution. We set a range ${\bf r}_{j}$ within which each phase can take values and introduce the probability distribution $p(\theta_j)$.
The average overlap $\bar{f}_N$ for the noisy linear
cluster state becomes
$\bar{f}_N = 2^{-N} \sum_{{\bf z_i}}\prod^{N-1}_{j=1}\{\int_{\vert{\bf r}_{j}\vert}
p(\theta_j) e^{i \theta_j}d \theta_j\}^{z_j z_{j+1}}$,
where $\vert{\bf r}_{j}\vert$ is the {\it width} of the range of variation of $\theta_{j}$ and $p(\theta_{j})$ depends on the specific physical model used in the cluster state generation. However, the nature of the fluctuations of the phases is characterized by the way in which the interactions among the elements of a lattice are realized and no {\it universal model} can be found. In Fig.~\ref{fig:fig2} {\bf (a)} we provide an example of $\bar{F}_{N}$ for a flat $p(\theta_{j})$ distribution, for the case of linear clusters. 
A large deviation from the ideal state is found and we have checked that this qualitative behavior holds
irrespective of the model for the unwanted phase distribution~\cite{noidopo}. This analysis can be extended to two-dimensional cluster states where analogous qualitative results can be found. 

The structure of the quantum correlations has been found to be profoundly different from what is found in $\ket{\phi}_{{\cal C}_N}$. Genuine multipartite entanglement is shared between the subparties of ideal cluster states, where the entanglement 
is encoded in the state as a whole \cite{rb0}. Any reduced bipartite state, obtained by tracing
all the qubits but an arbitrary pair, is separable as it does not violate the necessary and sufficient Peres-Horodecki criterion for separability of a mixed qubit
state~\cite{pereshorodecki}. However $S^{ab}_{D}$'s alter this result. For example, take a linear cluster of $N=3$, the ideal state is locally equivalent to a $GHZ$ state. However, two unwanted phases are embedded in the corresponding noisy state $\ket{\phi}_{{\cal C}_3}^D$.
The partial trace of $\ket{\phi}_{{\cal C}_3}^D\!\bra{\phi}$ over the third qubit gives a bipartite state
which violates the Peres-Horodecki criterion for
$\theta_{2}\neq{k\pi}$ ($k=0,1,..$) and
$\forall\,\theta_{1}\neq{\pi}$. This is a characteristic shared by
all the two-qubit states obtained by tracing qubits $3$ to $N$ in a $\ket{\phi}^{D}_{{\cal C}_N}$. 
In order to quantify the entanglement of the reduced density matrices $\rho_{ij}$ ($i,j\in[1,N]$), we use the {\it concurrence}, 
$C_{ij}=\max\{0,2\alpha^{ij}_{1}-\mbox{Tr}(\tilde{\rho}_{ij})\}$, where $\alpha^{ij}_{1}\ge\alpha^{ij}_{2}\ge{\alpha^{ij}_{3}}\ge{\alpha^{ij}_{4}}$ are the eigenvalues of ${\tilde{\rho}_{ij}}=\sqrt{\rho_{ij}(\sigma^{i}_{y}\otimes\sigma^{j}_{y})\rho^{*}_{ij}(\sigma^{i}_{y}\otimes\sigma^{j}_{y})}$ ($\rho^{*}_{ij}$ is the complex conjugate of $\rho_{ij}$ in the computational basis)~\cite{concurrence}.
We find that 
only nearest-neighbor bipartite entanglement is settled in a symmetric way along an arbitrarily long noisy cluster ({\it e.g.} ${C}_{12}={C}_{(N-1)N}$, irrespective of $N$), while
any non-nearest neighbor entanglement is absent due to randomness, more pronounced in pairs of non-nearest neighbor qubits.
Indeed the corresponding density matrices 
depend, in general, on more unwanted phases than those of nearest neighbor ones. 
At $\theta_{j}=\pi\,(\forall{j})$ no entanglement is found, as in  this case $S^{ij}_{D}\equiv\openone^{ij}$.  
We have explicitly considered the average concurrence obtained by assuming Gaussian fluctuations of each $\theta_{j}$, around $\theta_j=0$ and with a standard deviation $\sigma$. We find that as $\sigma$ increases (up to $\sigma=1$),
the fragile quantum correlations of the {\it bridging pairs} $(i,i+1)$ with $i\in[2,N-2]$ disappear, breaking the quantum channel connecting qubit $1$ to qubit $N$. This is due to the fact that the qubits in
these pairs are exposed to more randomness than those in the pairs $(1,2)$ and
$(N-1,N)$. We are left with the entangled mixed states of these {\it extremal} pairs which can mutually share only classical correlations. By increasing the randomness, even this entanglement will disappear, eventually. 
This analysis can be extended to the case of arbitrary $N$, despite 
the difficulties in finding the reduced density matrices of large cluster states. Thus, the multipartite entanglement is reduced to the benefit of
bipartite correlations which may be very fragile against
fluctuations in the unwanted phases, possibly leading to a complete
entanglement breaking effect. 

In addition to studying imperfect generation of cluster states, it is also necessary to understand effects of environmental decoherence. This is important for physical realizations, as the accuracy of a gate simulated on a cluster state interacting with an environment may be spoiled. We consider decoherence due to individual dephasing channels affecting each qubit in the cluster, a model which is relevant in practical situations of qubits exposed to locally fluctuating potentials. Here, the off-diagonal elements of a single-qubit density matrix $\rho^{d}$ decay as $e^{- \Gamma}$, with $\Gamma$ the rescaled dephasing time.
For a single qubit prepared in $|+ \rangle$, the fidelity is $F=\langle + |\rho^{d}| + \rangle=\frac{1}{2}(1+e^{- \Gamma })$. For larger registers, it is relevant to compare the behavior of cluster states under dephasing with other multiqubit states such as $GHZ$ and $W$ states. For $N$-qubit $GHZ$ and $W$ states ($N\ge3$), the state fidelities are $F^{GHZ}_N=(1+e^{- \Gamma N })/2$ and $F^{W}_N=(1+(N-1)e^{- 2 \Gamma })/N$, while for $N$-qubit ($N\ge2$) linear cluster states
$F^{C_{lin}}_N={2^{-N}}\sum_{h=0}^{N}B(N,h)e^{- \Gamma h }$,~
with the binomial coefficient $B(N,h)$. The expression for a $N\times{N}$ cluster is found for $N\rightarrow{N}^2$. These functions are plotted against $N$ and shown in Fig.~{\ref{fig:fig2}} {\bf (b)}, where one can see that when state fidelity is used as a benchmark, linear cluster states are very fragile against individual dephasing.

{\it Noisy cluster state QC}- To analyze the noise model in computational protocols, we briefly review the workings of the one-way QC. If a unitary operation $U_g$ has to be performed on an input state $|\psi_{\rm{in}} \rangle$ of $n$ logical qubits, the idea is to use a cluster in a particular physical configuration ${\cal{C}}(g)$. The input state is encoded in the $\sigma_z$ eigenbasis in a section ${\cal C}_{I}(g)$ of the cluster consisting of $n$ qubits. This section is then entangled to the rest of the cluster through the ideal entangling operations. A measurement pattern ${\cal{M}}^{(g)}$, {\it i.e.} a set of single-qubit measurements along particular directions in the Bloch sphere, is performed on all the qubits of the cluster except a part of it which embodies the output register ${\cal C}_{O}(g)$. 
The input and output logical state in the simulation of $g$ are related via $|\psi_{\text{\rm{out}}}\rangle = U_g U_\Sigma(\{s_i\})\,|\psi_{\text{\rm{in}}} \rangle$, where $U_\Sigma(\{s_i\})$ is a local {\it byproduct} operator which acts on the output logical qubits and depends on the set of outcomes $(\{s_i\})$ obtained after the measurements of ${\cal{M}}^{(g)}$~\cite{RBH}. 

It is not obvious how a particular QC protocol is affected by the exposure of a cluster state to a dephasing channel and it is the specific nature of the one-way QC which makes this point non-trivial. Considering that in order to perform a computation, the qubits in a cluster are measured and thus removed from the dynamics of the system, one would conjecture that in this way, they no longer contribute to the expected decrease of the gate fidelities due to dephasing~\cite{noidopo}. In the context of QC protocols with noisy cluster states, here we assess the effects of redundant qubits not functional to the specific simulation to be performed.
These qubits can be {\it removed} from the cluster configuration by measuring them in an appropriate basis. In the ideal case, if we measure in the $\sigma_x$ eigenbasis, we effectively obtain a reduced cluster state, where the measured qubits have no influence~\cite{RBH}. In general, the new cluster state will satisfy eigenvalue equations with a new set $\{ \kappa' \}$. For example, a five-qubit linear cluster (qubits labelled from $1$ to $5$) associated with the set $\{\kappa\}=\{0\}$ can be reduced to a three-qubit one, simply by measuring $3$ and $4$ in the $\sigma_x$ eigenbasis. This gives a cluster state $\ket{\phi_{ \{ \kappa' \}}}_{{\cal C}_3}$ with $\{\kappa'\}=\{0,s^x_4,s^x_3\}$. Here, $s^{x}_{i}=0\,(s^{x}_{i}=1)$ corresponds to qubits $i=3,4$ being in $\ket{+}$ $(\ket{-})$. On the other hand, measurements in the $\sigma_z$ eigenbasis {\it remove} a qubit from a cluster too, but also break any intra-cluster connection bridged by the qubit~\cite{RBH}. While this strategy does not affect gate performances in the ideal one-way QC, when the model in Eq.~(\ref{Sabdefdirty}) is contemplated, the removal of every redundant qubit determines the spread of the noise (in terms of random phases) from the removed qubits to the remainder of the cluster. We refer to this effect as the {\it inheritance} of noise by the survivor qubits. In turn, this results in additional spoiling mechanisms reducing the fidelity of the gate being simulated. The conclusions of our study do not qualitatively depend on the specific gate we consider. Thus, to give evidence of these effects, we go directly to the case of a ${\sf CNOT}$. 

\begin{figure}[b]
\hspace*{0.5cm}\hspace*{3.5cm}{\bf (c)}\\
\centerline{\psfig{figure=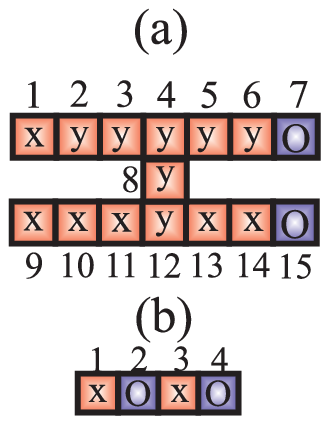,width=2.5cm,height=3.1cm}\hspace*{0.8cm}\psfig{figure=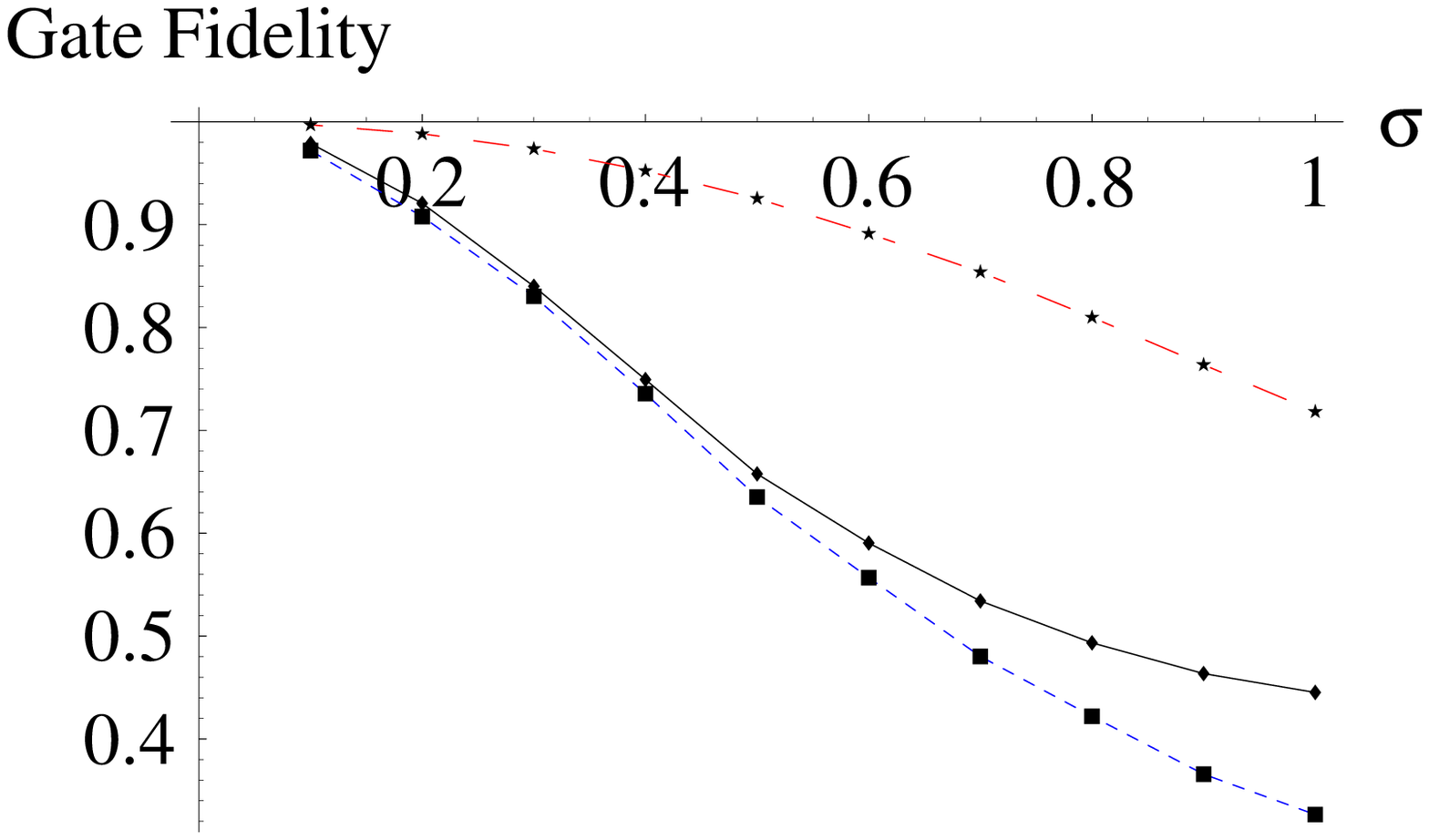,width=5.0cm,height=3.3cm}}
\caption{{\bf (a)}: The \textit{squashed-}I configuration~\cite{RBH}.
{\bf (b)}: Scheme for a $\sf CNOT$ with a four-qubit cluster. {\bf (c)}: Average fidelity of a $\sf CNOT$ against the $\sigma$ of Gaussian distributions for each $\theta_{j}$. From top to bottom we show the results for panel ${\bf (b)},\,\bf (a)$ and the squashed-I with an additional bridge-qubit respectively.}
\label{cnotBBB}
\end{figure}

In order to perform a {\sf CNOT} between a control qubit $|c_{\rm in} \rangle=a |0 \rangle + b |1 \rangle$ and a target qubit $|t_{\rm in} \rangle=c |0 \rangle + d |1 \rangle$, the scheme in~\cite{RBH} uses $15$ qubits, where ${\cal{M}}^{({\sf CNOT})}$ consists of measurements in the $\sigma_{x,y}$ eigenbases as shown in Fig.~\ref{cnotBBB} {\bf (a)}. After the measurements, {\it decoding} operators $\tilde{U}_{\Sigma}^{c,t\dag}(\{s_i\})= \sigma_z^{\gamma_z^{c,t}}\otimes\sigma_x^{\gamma_x^{c,t}}$ are applied to qubits $7$ and $15$ (which embody the logical control and target), where $\gamma_{x,z}^{c,t}$ depend on the outcomes $s_{i}$ of the individual measurements~\cite{RBH}.
A $\ket{\phi}^{D}_{{\cal C}_{15}}$ state is needed in the noisy gate simulation and to construct it, we take small subclusters which are then mutually entangled. For instance, the subclusters $\ket{\phi}^D_{1,2,3,4}$ and $\ket{\phi}^D_{5,6,7}$ are entangled as $S^{4,5}_{D}\ket{\phi}^D_{1,2,3,4}\ket{\phi}^D_{5,6,7}=\ket{\psi}^D_{1,2,3,4}\ket{\phi}^D_{5,6,7}+\ket{\chi}^D_{1,2,3,4}\ket{\eta}^D_{5,6,7}-e^{i\theta_{4}}\ket{\chi}^D_{1,2,3,4}\ket{\mu}^D_{5,6,7}$. Here, 
$\ket{\psi}^D$ ($\ket{\eta}^D$) is the part of a subcluster state with the last (first) qubit in ${\ket{0}}$, while $\ket{\chi}^D$ ($\ket{\mu}^D$) is the part with the last (first) qubit in $\ket{1}$ and the labels of the qubits are explicitly indicated. By applying the pattern shown in Fig.~\ref{cnotBBB} {\bf (a)}, the gate fidelity can be evaluated.
To illustrate the effect of imperfections, we address the case where qubits measured in the $\sigma_{x}~(\sigma_{y})$ eigenbasis are all found in $\ket{+}$ ($\ket{+}_{y}\propto\ket{0}+i\ket{1}$) 
and the required decoding operator is $\tilde{U}_{\Sigma}^{\dag}(\{0\})=\sigma_{z}^{(7)}\otimes\openone^{(15)}$. To take into account the phase randomness, we have averaged the gate fidelity over Gaussian distributions centered on $\theta_{j}=0$ and with a standard deviation $\sigma$ for each phase in $\ket{\phi}^{D}_{{\cal C}_{15}}$. The results for $\sigma\in[0.1,1]$ are shown in Fig.~\ref{cnotBBB} {\bf (c)} (solid curve) where the decay of the fidelity against the increasing randomness is evident. The input states are normalized
and we take $a=c=0.5$. Similar behaviors are observed for other choices of $a$ and $c$. To see {\it noise inheritance} effects, we modify the configuration of Fig. \ref{cnotBBB} {\bf (a)} by adding a bridging qubit between $8$ and $12$, thus considering a state $\ket{\phi}^{D}_{{\cal C}_{16}}$. This qubit is redundant and is {\it removed} via a $\sigma_{x}$-measurement. The $15$-qubit cluster state is retrieved via local operations on qubit $12$ (the removal is equivalent to a {Hadamard} gate between qubits $16$ and $12$). When the gate fidelity is calculated, a faster decrease against $\sigma$ is observed (Fig. \ref{cnotBBB} {\bf (c)}, bottom curve). The same effect is found if the information flow through a linear cluster state is studied~\cite{RBH}. In this case, the longer the noisy linear cluster across which information is transferred (thus requiring many measurements), the worse the transfer fidelity \cite{noidopo}. This is important evidence of the effect of measurements on noisy cluster states. 
The elimination of redundant qubits and the measurement pattern spread the $\theta_{j}$ in the noisy cluster state, affecting the QIP protocol. 
If purification procedures are not considered (usually {\it expensive} in terms of resources), a key method in counteracting both the noise models we have considered is the use of small cluster states. 

Indeed, the drop in fidelity seen for the {\it squashed-}I cluster and our discussion about information flow legitimize some doubts about the convenience of many-qubit configurations in the one-way QC and stimulate research for more compact protocols, in order to bypass noise inheritance or the effect of dephasing in the dynamics of a cluster. The scheme for a $\sf CNOT$ in Fig.~\ref{cnotBBB} {\bf (b)} proves the existence of such economical configurations. It consists of a four-qubit linear cluster and a measurement pattern made by the $\sigma_{x}$ measurement of qubits $1$ and $3$ (encoding the input target and control state). In order to demonstrate the gate simulation, we assume perfect entangling operations $S^{ab}$, so that the four-qubit cluster state is $\ket{\psi}_{1,2}\ket{\eta}_{3,4}+\ket{\psi}_{1,2}\ket{\mu}_{3,4}+\ket{\chi}_{1,2}\ket{\eta}_{3,4}-\ket{\chi}_{1,2}\ket{\mu}_{3,4}$ with $\ket{\psi}_{1,2}\propto(a\ket{00}+b\ket{10})_{1,2},\ket{\chi}_{1,2}\propto({a}\ket{01}_{1,2}-b\ket{11})_{1,2},\ket{\eta}_{3,4}\propto{c}(\ket{00}+\ket{01})_{3,4}$ and $\ket{\mu}_{3,4}\propto{d}(\ket{10}-\ket{11})_{3,4}$. Measuring qubits $1$ and $3$ onto $\ket{\pm}_{1,3}$, qubits $2$ and $4$ are left in a state locally equivalent to the output state of a ${\sf CNOT}_{c=4,t=2}$ (in the $\sigma_{x}$ eigenbasis)~\footnote{The decoding operators are $\tilde{U}^{\dag}_{\Sigma}(s_{1},s_{3})=[\sigma^{(2)}_{x}]^{s_1}\otimes[\sigma^{(4)}_{x}]^{s_1\oplus{s}_3}$, with $\oplus$ the logical XOR.
}. When $S^{ab}\rightarrow{S}^{ab}_{D}$, the cluster state becomes noisy and the gate fidelity is affected by the distributions for $\theta_{j}$. However, the small number of qubits limits any spoiling effect and improves the average gate fidelity (Fig. \ref{cnotBBB} {\bf (c)}, top curve). 
This reinforces the idea that more economical schemes for cluster state QC exist, where the number of qubits involved corresponds to the number of parameters in the input state. In fact, single-qubit gates can be done via a two-qubit cluster and one measurement along an appropriate direction~\cite{noidopo}. This circumstantial evidence suggests that our scheme uses the smallest  qubit register needed for a two-qubit gate and is thus optimal.

The four-qubit {\sf CNOT} can be realized in an optical setup, requiring two pure entangled states (encoded in photonic polarizations) and an entangling gate. A pure state of arbitrary entanglement can be produced using the entanglement between two field modes generated by concatenating Type-$I$ parametric-down-conversion (PDC) processes~\cite{kwiat}. In this scheme, the polarization of the pump field 
sets the entanglement at the output of the PDC process and encodes arbitrary target and control input states in pairs of output modes ({\it i.e.} pairs $1+2$ and $3+4$) without postselection. By adapting the scheme~\cite{sara}, modes $2$ and $3$ can be entangled through an effective controlled $\pi$-phase gate. This protocol results in the four-qubit cluster state we have addressed~\cite{noidopo}. 
 
{\it Remarks}- We have shown that realistic imperfections in the generation and processing of cluster states affects the model for one-way QC.
To counteract these effects, the dimension of a cluster has to be minimized. In this context, we have demonstrated an experimentally realizable four-qubit {\sf CNOT} which uses the minimum number of qubits for a cluster state based {\sf CNOT}. Our proposal demonstrates the possibility of designing more compact schemes for cluster state QIP. This theoretically challenges the way in which cluster state-QC has been thought about so far and allows for more controllable experimental  implementations.

We acknowledge support by EPSRC, DEL and IRCEP. MST thanks the E. Schr\"{o}dinger Institute.


\end{document}